\documentclass[a4paper]{amsart}

\usepackage{graphicx}
\usepackage[dvipsnames]{xcolor}
\usepackage{epstopdf, epsfig}
\usepackage{amsaddr}
\usepackage{amsmath,amsfonts,amssymb,amsgen,amsbsy,dsfont}
\usepackage{subfig}
\usepackage{caption}
\usepackage{hyperref}

\newcommand{\depth}{d}                              
\newcommand{\sur}[1]{{#1}_\mathrm{s}}                    
\renewcommand{\bot}[1]{{#1}_\mathrm{b}}                  
\newcommand{\eqdef}{\stackrel{\text{\tiny{def}}}{=}} 

\newcommand{\rf}[1]{(\ref{#1})}

\newcommand{\ud}{\mathrm{d}}

\newcommand{\ue}{\mathrm{e}}
\newcommand{\ui}{\mathrm{i}}
\newcommand{\upi}{\pi}

\renewcommand{\Re}{\operatorname{Re}}
\renewcommand{\Im}{\operatorname{Im}}

\newcommand{\half}{{\textstyle{\frac{1}{2}}}}

\newcommand{\sixth}{{\textstyle{\frac{1}{6}}}}

\newcommand{\cs}{c_2}
\newcommand{\ce}{c_1}

\begin{document}

\title[Recovery of rotational waves]{Recovery of steady rotational wave profiles from pressure measurements at the bed}

\author[D. Clamond]{Didier Clamond}
\address{Universit\'e C\^ote d'Azur, CNRS UMR 7351,  Laboratoire J. A. Dieudonn\'e, 
Parc Valrose, 06108 Nice cedex 2, France}
\email{didier.clamond@univ-cotedazur.fr}

\author[J. Labarbe]{Joris Labarbe}
\address{Universit\'e C\^ote d'Azur, CNRS UMR 7351,  Laboratoire J. A. Dieudonn\'e, 
Parc Valrose, 06108 Nice cedex 2, France}

\author[D. Henry]{David Henry}
\address{School of Mathematical Sciences, University College Cork, Cork, Ireland}

\subjclass[]{}

\begin{abstract}
We derive equations relating the pressure at a flat seabed 
and the free-surface profile for steady gravity waves 
with constant vorticity. The resulting set of nonlinear 
equations enables the recovery of the free surface from 
pressure measurements at the bed. Furthermore, the flow 
vorticity is determined solely from the bottom pressure as 
part of the recovery method. This approach is applicable 
even in the presence of stagnation points and its efficiency 
is illustrated via numerical examples. 
\end{abstract}

\maketitle

\section{Introduction}

In this paper, we present a formulation of the rotational water 
wave problem which enables the recovery of nonlinear surface 
gravity wave profiles from pressure measurements at the seabed, 
for steady flows with constant vorticity.
The determination of the wave profile is achieved by numerically 
solving a set of nonlinear equations, with our inverse recovery 
procedure having the significant side-benefit of also determining 
the vorticity $\omega$ directly from bottom pressure measurements. 
The presence of vorticity greatly complicates the mathematical 
problem, and  the recovery of fully nonlinear rotational water wave 
profiles from pressure measurements has hitherto  proven unattainable 
(although explicit surface-profile recovery formulae for linear, and 
weakly nonlinear, rotational water waves were derived by 
\cite{HenryThomas2018} for arbitrary vorticity distributions). 

The reconstruction of water wave surface profiles from bottom pressure 
measurements is a theoretically challenging issue with important 
applications in marine engineering. Measuring the surface of water 
waves directly is  difficult and costly, particularly in the ocean, 
so a commonly employed alternative is to calculate the free-surface 
profile of water waves using measurements from submerged pressure 
transducers. To do so requires the construction of either a suitable 
pressure-transfer function (for linear waves), or a surface 
reconstruction procedure (for nonlinear waves), the determination 
of which corresponds to a difficult mathematical problem.

Until quite recently, most  surface reconstruction formulae were 
applicable only to the restricted setting of linear water waves, 
and even then solely for irrotational flows. 
First approaches towards surface reconstruction formulae for 
nonlinear irrotational waves appeared in \cite{Constantin2012,OVDH}, 
however these formulae are quite involved. Exact tractable relations 
were derived in \cite{Clamond2013,ClamondConstantin2013} which 
permit a straightforward numerical procedure for deriving the free 
surface from the bottom pressure.  A significant advantage of these 
approaches is that they work directly with nonlinear waves in the 
physical plane, allowing recovery of nonlinear wave-profiles up to, 
and including, Stokes wave of greatest height \cite{ClamondHenry2020}.
The robustness of this nonlinear wave surface reconstruction approach 
is further illustrated in this paper by expanding it to encompass 
flows with constant vorticity.

Incorporating vorticity in the water wave problem is vital for 
capturing fundamental physical processes relating to wave-current 
interactions \cite{ThomasKlopman1997}, however it significantly 
complicates all theoretical considerations \cite{Constantin2011}. 
We note that while it was rigorously proven by \cite{Henry2013} 
that the profile-recovery problem is well-posed for nonlinear {\em 
solitary} waves with arbitrary (real analytic) vorticity distributions, 
it  remains an open question whether the inverse recovery problem is 
well-posed for {\em periodic} waves.
 Even in the simplified setting of constant vorticity, the water wave 
 problem exhibits features not encountered in the irrotational case.  
 In particular, we note such flows may contain stagnation points (and 
 critical layers) in the fluid interior, and waves which possess 
 overhanging profiles. The possibility  of overhanging waves was 
 first observed numerically 
 \cite{DaSilvaPeregrine1988,OkamotoShoji2001} and their possible 
 existence was recently rigorously proven 
 \cite{ConstantinVarvaruca2011,ConstantinStraussVarvaruca2016}. 
 We note that the surface recovery approach introduced in this paper 
 is applicable to flows containing stagnation points, however we 
 must exclude overhanging profiles {\em a priori}. 
 It is assumed throughout that the surface profile is a graph 
 (overhanging waves cannot occur for `downstream' waves, c.f. 
 \cite{ConstantinStraussVarvaruca2021}). 

\section{Preliminaries}
\label{secmathdef}

In the frame of reference moving with a travelling wave of permanent shape, the flow beneath the wave reduces 
to a steady motion with respect to the moving coordinate system. Thus, the wave phase velocity $c$ is constant 
in any Galilean frame of reference. 
Let $(x,y)$ be a Cartesian coordinate system moving with the wave, $x$ being the horizontal coordinate and $y$ 
the upward vertical coordinate. 
We define accordingly the fluid domain as $\Omega = \{(x,y):x\in\mathds{R},-\depth\leqslant y\leqslant\eta(x)\}$, 
where $y=-\depth$ and $y=\eta(x)$ correspond, respectively, to the solid bottom and the free-surface (both 
impermeable). In addition, $\left(u(x,y),v(x,y)\right)$ denotes the velocity field in the moving frame.
We assume the wave is $L=(2\upi/k$)-periodic (with $k\to0$ for solitary waves) in the $x$-direction, and we 
denote by $y=0$ the mean water level. The latter equation expresses the fact that $\left<\eta\right>=0$, 
where $\left<\cdot\right>$ is the Eulerian average operator over one period, that is, 
\begin{equation} 
\label{defmean}
\left< \eta \right> \eqdef \frac{k}{2\upi} \int_{-\upi/k}^{\upi/k} \eta(x) \ud x = 0.
\end{equation}
The flow is governed by the balance between the restoring gravity force and the inertia of the system. 
With constant density $\rho>0$, the equation of mass conservation and Euler equations (defined in $\Omega$) 
are, respectively, 
%
%
%
\refstepcounter{equation}
\[
\partial_x u + \partial_y v = 0, \quad
u\/\partial_x u + v\/\partial_y u = -\partial_x P/\rho, \quad
u\/\partial_x v + v\/\partial_y v = -\partial_y P/\rho - g, 
\eqno{(\theequation{\/\mathit{a},\mathit{b},\mathit{c}})}\label{ee}
\]
where $P(x,y)$ denotes the pressure. 
As a general notation, subscripts `{\rm b}' denote all quantities written at the bed $y=-\depth$ whereas subscripts 
`{\rm s}' denote all quantities written at the free surface $y = \eta(x)$. The effect of surface tension being 
neglected, on the free surface we must have  the dynamic boundary condition
\begin{equation} \label{d}
\sur{P} = P_\text{atm} ,
\end{equation}
where $P_\text{atm}$ is the (constant) atmospheric pressure. 
The free surface and the rigid bed are impermeable interfaces, giving the kinematic boundary conditions 
(with $\eta_x\eqdef\ud\eta(x)/\ud x$)
\refstepcounter{equation}
\[
\sur{v} = \sur{u} \eta_x, \qquad \bot{v} = 0,
\eqno{(\theequation{\/\mathit{a},\mathit{b}})}\label{kt}
\]
respectively, while the rotational character of the flow is ensured by requiring
\begin{equation}
\label{i}
\partial_x v - \partial_y u \eqdef \omega \quad (= \text{constant}).
\end{equation}
Equations (\ref{ee})--(\ref{i}) are the governing equations for rotational (of 
constant vorticity $\omega$) travelling water waves in a frame of reference moving with 
the wave. 

For incompressible flows where  (\ref{ee}{\it a}) holds we can define a streamfunction $\psi$ such that $u=\partial_y\psi$ and $v=-\partial_x\psi$. As the flow is steady 
and the free-surface is impermeable, it follows that the free-surface is a streamline, that is, the streamfunction  is constant $\psi=\sur{\psi}$ at the free-surface (similarly, the streamfunction is constant  $\psi=\bot{\psi}$ at the bed). 

Equations (\ref{ee}) can  be integrated to 
\begin{equation}
\label{bernbase}
2 p + 2 g y + u^2 + v^2 = \sur{B} - 2\omega (\psi - \sur{\psi}),
\end{equation}
for some constant $\sur{B}$, where $p(x,y)\eqdef[P(x,y)-P_\text{atm}]/\rho$ is  
a normalised relative pressure. 
%
Equation \eqref{bernbase} is a Bernoulli equation, and we note that the Bernoulli 
integral $B(\psi) \eqdef \sur{B} - 2\omega(\psi - \sur{\psi})
$ is constant for an irrotational motion (i.e., when $\omega=0$).

\section{Definition of the parameters}

From the definition (\ref{defmean}) of the mean water level and by averaging expression (\ref{bernbase}) written at the free surface, we obtain a definition for the constant $\sur{B}$ in the form
\begin{equation}
\label{defBsur}
\sur{B} = \left< \sur{u}^{\,2} + \sur{v}^{\,2} \right>.
\end{equation}
As the frame of reference moving with the wave is Galilean, there is no mean acceleration. 
For steady waves with constant vorticity, the zero-mean horizontal acceleration condition is perforce
satisfied, but the condition for zero mean vertical acceleration yields
\begin{align}
0\ &= \left< \int_{-\depth}^\eta \left[ u \frac{\partial v}{\partial x}
+ v \frac{\partial v}{\partial y} \right] \ud y \right> =
\left< \int_{-\depth}^\eta \frac{\partial}{\partial y} \left[ \frac{u^2+v^2}{2}
+ \omega \psi \right] \ud y \right> \nonumber \\
\label{mva0}
&= \half \left< \sur{u}^{2} + \sur{v}^{2} \right> - 
\half \left< \bot{u}^{2} \right>
+ \omega\,(\sur{\psi} - \bot{\psi}) .
\end{align}
This furnishes at once an alternative relation for the Bernoulli constant
\begin{equation} 
\label{defBbot}
\sur{B} = \left< \bot{u}^{2} \right> + 2\omega\,(\bot{\psi} - \sur{\psi}) ,
\end{equation}
and, since $\sur{p}=0$, relation \eqref{mva0} implies the average pressure at the bottom is 
\begin{align}
\left< \bot{p} \right> &= - \left< \int_{-\depth}^\eta \frac{\partial p}
{\partial y} \ud y \right> = \left< \int_{-\depth}^\eta \left[ u
\frac{\partial v}{\partial x} + v \frac{\partial v}{\partial y} +
g\right] \ud y\right> = g \depth. 
\label{defdepth}
\end{align}
Relation \eqref{defdepth} provides a mechanism for determining the mean water depth $\depth$ from bottom pressure measurements.
For later convenience, define the alternative Bernoulli constant
\begin{equation}
\label{defBi}
\bot{B} \eqdef \sur{B} - 2\omega\, (\bot{\psi} - \sur{\psi}) =
\left< \bot{u}^{\,2} \right>, 
\end{equation}
where, as expected, both Bernoulli constants $\bot{B}$ and $\sur{B}$ coincide for irrotational flows. 
The vorticity $\omega$ being constant, exploiting the free surface impermeability gives 
\begin{align}
\omega\depth &= \left< \int_{-\depth}^\eta \left[ \frac{\partial v}
{\partial x} - \frac{\partial u}{\partial y} \right] \ud y \right> = 
- \left< \eta_x \sur{v} \right> - \left< \sur{u} \right> + 
\left< \bot{u} \right> \nonumber \\
&= \left< \bot{u} \right> - \left< \left(1 + \eta_x^{2} \right) \sur{u} \right>.
\label{eqwuvce}
\end{align}
Expression \eqref{eqwuvce}  provides a means of determining the vorticity $\omega$ 
in terms of $\bot{u}$, $\sur{u}$, $\depth$ and $\eta$. 
In the same vein, a relation not involving velocity evaluation along the flat bed is given by
\begin{align}
\frac{\omega}{2} \left<h^2\right> &= \left< \int_{-\depth}^\eta \left[ \frac{\partial v}
{\partial x} - \frac{\partial u}{\partial y} \right] (y+d) \ud y \right> =  
- \left< h \eta_x \sur{v} \right> - \left< h \sur{u} \right> + \sur{\psi}- 
\bot{\psi} \nonumber \\
&= \sur{\psi} - \bot{\psi} - \left< \left(1+\eta_x^{2}\right) h \sur{u}\right>,
\label{eqw}
\end{align}
where $h(x)\eqdef\eta(x)+\depth$ is the total water depth. Together with \eqref{defBi}, relation \eqref{eqw} can be expressed
\begin{equation}
\label{BsBb}
\sur{B} = \bot{B} - \omega^2 \left< h^2 \right> - 
2 \omega \left< \left(1+\eta_x^{2}\right) h \sur{u} \right>.    
\end{equation}

As for irrotational motions, Stokes' first and second definitions of the 
phase celerities \cite{Clamond2017,KishidaSobey1988} can be applied, resulting 
in the expressions
\begin{align}
\ce &\eqdef - \left< \bot{u} \right>
= -\omega\depth - \left< \left(1+\eta_x^{2}\right) \sur{u} \right>, 
\label{defce} \\
\cs &\eqdef - \left< \frac{1}{\depth} \int_{-\depth}^\eta u \ud y \right>
= \frac{\bot{\psi} - \sur{\psi}}{\depth} = - \frac{\omega\depth}{2} - 
\frac{\omega\left<\eta^2\right>}{2\depth} - \frac{\left< \left(1+\eta_x^{2}\right) h \sur{u} \right>}{\depth}.
\label{defcs}
\end{align}
Here $\ce$ and $\cs$ are the wave speeds observed in frames of reference without 
mean horizontal velocity at the bed, and without mean flow, respectively. 
In the irrotational case ($\omega=0$), it can be shown 
\cite{ClamondDutykh2018b} 
that $\cs\to\ce$ and $\bot{B}=\sur{B}\to\ce^2$ as $\depth\to\infty$ or as 
$k\to0$, but $\cs\approx \ce$ in the linear wave regime. For constant 
vorticity $\omega\neq0$, matters are more complex, even 
at the linear level: $\cs \approx \ce-\omega d/2 \not\approx \ce$, while 
$\ce\approx c^\pm_0$ where the linear phase speed  \cite{BrinkKjaer1976,KishidaSobey1988}
\begin{equation}
c^\pm_0=-\omega d+\half k^{-1}\omega\tanh(k d)\pm\half k^{-1}\sqrt{\omega^2
\tanh(k d)^2+4gk\tanh(k d)},
\end{equation}
solves a linear dispersion relation with symmetry property $c_0^{\pm}(-\omega)
=-c_0^{\mp}(\omega)$. Hence, without loss of generality in subsequent
considerations, we assume that $\ce>0$ (that is, the wave propagates toward the increasing $x$-direction 
in the frame of reference without mean velocity at the seabed) and allow the vorticity 
to take either sign.

\section{Holomorphic functions}

When $\omega=0$ the flow is irrotational and one can use the powerful theory of holomorphic 
functions \cite{Clamond2013,ClamondConstantin2013,ClamondHenry2020}. In the case where 
$\omega\neq0$ one can still use this technique following Helmholtz representation 
\cite{Aris1962}. Thus, introducing $U$, $V$ and $\Psi$ (functions of both $x$ and $y$) as  
%
\refstepcounter{equation}
\[ 
U \eqdef u + \omega\,(y+\depth), \qquad V \eqdef v, \qquad 
\Psi \eqdef \psi + \half\/\omega\,(y+\depth)^2,
\eqno{(\theequation{\/\mathit{a},\mathit{b},\mathit{c}})}\label{defUV}
\]
from which, using (\ref{i}), straightforward calculations show that $U = \partial_y 
\Psi$, $V = -\partial_x \Psi$ and  $\partial_x V = \partial_y U$, implying that the 
velocity field $(U,V)$ is curl-free. At the bed $y=-\depth$, we have
\refstepcounter{equation}
\[ 
\bot{U} = \bot{u}, \qquad \bot{V} = \bot{v} = 0, \qquad 
\bot{\Psi} = \bot{\psi},
\eqno{(\theequation{\/\mathit{a},\mathit{b},\mathit{c}})}\label{defUVbot}
\]
while at the free surface $y=\eta(x)$
\refstepcounter{equation}
\[ 
\sur{U} = \sur{u} + \omega h, \qquad \sur{V} = \sur{v}, 
\qquad \sur{\Psi} = \sur{\psi} + \half\/\omega\/h^2.
\eqno{(\theequation{\/\mathit{a},\mathit{b},\mathit{c}})}\label{defUVsur}
\]
Note that $\sur{\Psi}$ is not uniform (unlike $\sur{\psi}$) while $\bot{\Psi}$ is a constant. 
Thus, introducing a velocity potential $\Phi$ such that $U = \partial_x\Phi$ and $V = \partial_y\Phi$, 
the complex potential and velocity are defined as
\refstepcounter{equation}
\[ 
F(z) \eqdef \Phi + \ui \Psi, \qquad W(z) \eqdef U - \ui V = \ud F/\ud z,
\eqno{(\theequation{\/\mathit{a},\mathit{b})}}
\]
where $z \eqdef x + \ui y$ is a complex coordinate in $\Omega$.
With these new dependent variables, the Bernoulli equation (\ref{bernbase}) evaluated at the free surface becomes
\begin{equation} 
\label{bernsur}
\sur{U}^{\,2} + \sur{V}^{\,2} + 2 \left( g\eta - \omega h \sur{U} \right) + (\omega h)^2 = \sur{B},
\end{equation} 
while at the bed it yields
\begin{equation} 
\label{bernbed}
2 \left( \bot{p} - g\depth \right) + \bot{U}^{\,2} = \sur{B} - 2\omega\,
(\bot{\psi} - \sur{\psi})  = \left< \bot{U}^{\,2} \right> = \bot{B}.
\end{equation}
Note that equations \eqref{bernsur} and \eqref{bernbed} can be rewritten, respectively,  
\refstepcounter{equation}
\[
\sur{U} = \omega h - \sqrt{(\sur{B}-2g\eta)/(1 + \eta_x^{\,2})},
\qquad
\bot{U} = - \sqrt{ \bot{B} - 2 \left( \bot{p} - g\depth \right)}, 
\eqno{(\theequation{\mathit{a},\mathit{b}})}\label{UsUbeta}
\]
the minus sign in front of the radicals being consequence of the choice $\ce>0$.
Since $\bot{V} = 0$, the complex velocity can then be expressed as
\begin{equation}
\label{Wz}
W(z) = - \sqrt{ \bot{B} - 2\bot{p}(z+\ui\depth) + 2g\depth },
\end{equation}
a relation which suggests the introduction of a complex pressure. Following 
\cite{ClamondConstantin2013}, we introduce a `complex pressure' function 
$\mathfrak{P}$ defined as 
\begin{align}
\label{P}
\mathfrak{P}(z) \eqdef &\,g\depth + \omega\,(\sur{\psi}-\bot{\psi})
- \half\left(W^2-\sur{B}\right),
\end{align}
which is holomorphic in the fluid domain $\Omega$. Note that the expression in \rf{P} is purely 
real when restricted to the flat bed, with $\bot{\mathfrak{P}} = \bot{p}$ on $y=-\depth$. 
Accordingly, $\bot{p}$ determines $\mathfrak{P}$ uniquely within the entire fluid domain $\Omega$, 
and so ${\mathfrak P}(z)=\bot{p}(z+\ui\depth)$. 
We note that since $p$ is not a harmonic function in the fluid domain \cite{Constantin2011}, 
it can coincide with the real part of $\mathfrak{P}$ only at $y=-\depth$.  

Similarly, as for irrotational waves in  \cite{Clamond2013,Clamond2018}, it is useful to introduce the holomorphic function $\mathfrak{Q}$ 
\begin{equation} 
\label{defQfun}
\mathfrak{Q}(z) \eqdef \int_{z_0}^z \left[ \mathfrak{P}(z') - g\depth \right] \ud z'
 = \frac{1}{2} \int_{z_0}^z \left[ \bot{B} - W(z')^2 \right] \ud z' , 
\end{equation}
where $z_0\in\Omega$ is an arbitrary constant.

\section{Equations for the surface recovery}

Integrating \rf{defQfun} along the free surface path, with the origin located 
at the crest (i.e. $z_0=\ui a$, $a\eqdef \eta(0)$ being the wave 
amplitude), one gets
\begin{equation} \label{Qs}
\sur{\mathfrak{Q}}(x) = \int_0^x \left[\/\sur{\mathfrak{P}}(x') - 
g\depth\/\right]\left(1 + \ui\eta_x'\right) \ud x', 
\end{equation}
where $\sur{\mathfrak{P}}(x) = g\depth  +(\bot{B} - \sur{U}^2 + \sur{V}^2)/2 
+ \ui\sur{U}\sur{V}$. With (\ref{kt}{\it a}), (\ref{defUVsur}{\it a,b}), 
(\ref{UsUbeta}{\it a}) and splitting real and imaginary parts, \rf{Qs} yields 
after some algebra
\begin{align} 
\label{Qsfull}
\sur{\mathfrak{Q}}(x) = &\int_{0}^{x} \left[ \Re{\left\{\sur{\mathfrak{P}}(x')\right\}} - g\depth - \Im{\left\{\sur{\mathfrak{P}}(x')\right\}} \eta_x' \right] \ud x' \nonumber \\ 
&+\ui\/(\eta-a)\left[ \hat{B} - \half(g+\omega^2d) (\eta+a) - \sixth\omega^2\left(\eta^2+a\eta+a^2\right)\right] ,
\end{align}
where $\hat{B}\eqdef(\sur{B}+\bot{B}-\omega^2d^2)/2$ is an expression involving Bernoulli constants at the surface and the bottom.
The imaginary part of \rf{Qsfull} provides an implicit relation for the surface elevation expressed in terms of the holomorphic function $\sur{\mathfrak{Q}}$
\begin{equation} 
\label{Qseta}
\Im\{\sur{\mathfrak{Q}}\} =  (\eta-a)\left[ \hat{B} - \half(g+\omega^2d) (\eta+a) - \sixth\omega^2\left(\eta^2+a\eta+a^2\right)\right].
\end{equation}
From the differentiation of \rf{Qs} and \rf{Qseta}, one gets 
the differential equation
\begin{equation}
\label{deta}
\frac{\ud\eta}{\ud\/x}\ = \frac{\Im\!\left\{\sur{\mathfrak{P}}\right\}}{\hat{B} 
- (g+\omega^2\depth) \eta - \half \omega^2 \eta^2 - \Re\!\left\{\sur{\mathfrak{P}}
\right\} + g\depth}.
\end{equation}
In the special case of irrotational motion ($\omega=0$), the differential 
equation  derived in \cite{ClamondConstantin2013} is recovered. 
Evaluating \rf{Qseta} at the trough ($x=-\pi/k$), bearing in mind relation 
\rf{BsBb}, one obtains an expression for $\hat{B}$ as
\begin{align} 
\label{Bhat}
\hat{B} &= \sur{B} + \half\omega^2 \left< \eta^2 \right> - \omega\left< h \sqrt{(\sur{B}-2g\eta)(1+\eta_x^2)} \right> , \nonumber \\
&= \half (g+\omega^2d) (a-b) + \sixth \omega^2 (a^2-ab+b^2) - (a+b)^{-1} \Im\{\sur{\mathfrak{Q}}(-\pi/k)\} ,
\end{align}
where $b\eqdef-\eta(-\pi/k)$ denotes the trough height (thus $H\eqdef a+b$ is the 
total wave height).

We now have algebraic expressions for the recovery of $\eta$ and $\hat{B}$, 
as functions of the remaining unknown parameters $a$, $b$ and $\omega$. 
Three relations are then needed to close our set of equations. These 
relations are obtained considering $\sur{\mathfrak{P}}=g\depth + 
\omega\,(\sur{\psi}-\bot{\psi})- \half\left(\sur{W}^{\,2}-\sur{B}\right)$ 
at the crest, at the trough and at an intermediate point. 
Using the surface impermeability with the decomposition \rf{defUV}, 
the squared complex velocity reduces to
\begin{align} \label{Ws}
\sur{W}^{\,2} &\eqdef (\sur{U} - \ui\sur{V})^2 = \omega^2 h^2 + 2\omega h 
\sur{u}\,(1 - \ui\eta_x) + \sur{u}^{\,2}\, (1 - \ui\eta_x)^2 , \nonumber \\
&= \omega^2 h^2 +\left[ (\sur{B}-2g\eta)/(1 + \ui\eta_x) - 2\omega h 
\sqrt{(\sur{B}-2g\eta)/(1+\eta_x^2)} \right]  (1 - \ui\eta_x) .
\end{align}
Following \cite{ClamondConstantin2013}, we substitute \rf{Ws} in the definition 
of $\sur{\mathfrak{P}}$ at the crest and the trough --- together with \rf{BsBb} 
and \rf{defcs} --- to get the two relations
\begin{align}\label{Pa}
\sur{\mathfrak{P}}(0) &= g h_c + \omega^2 \frac{\langle h^2 \rangle - h_c^2}{2} 
- \omega \left[ \left< h \sqrt{(\sur{B}-2g\eta)(1+\eta_x^{\,2})} \right> - h_c 
\sqrt{\sur{B} - 2ga} \right] , \\
\label{Pb}
\sur{\mathfrak{P}}\!\left(-\frac{\pi}{k}\right) &= g h_t + \omega^2 \frac{\langle 
h^2 \rangle - h_t^2}{2} - \omega \left[ \left< h \sqrt{(\sur{B}-2g\eta)(1+\eta_x^{\,2})} 
\right> - h_t \sqrt{\sur{B} + 2gb} \right],
\end{align}
where $h_t\eqdef d-b$ and $h_c\eqdef d+a$ are local depths under, respectively, 
the trough and the crest. As expected, both expressions reduce in the irrotational 
limit to the formulae derived in \cite{Clamond2013}. With \eqref{Pa} and \eqref{Pb} 
we have two relations that close the problem if $\omega$ is known. 
However, in practice $\omega$ is generally unknown {\em a priori\/}, so 
another independent relation must be introduced. 

A last relation is obtained considering the complex pressure 
$\sur{\mathfrak{P}}$ at an abscissa $x_0$ strictly between crest and 
trough. This point is chosen at a coordinate of median bottom pressure 
measurement such that  $\bot{p}(x_0)=(\max\bot{p}-\min\bot{p})/2$ to 
ensure a large distance from the crest and the trough. $x_0$ being thus 
chosen, $\Im\{\sur{\mathfrak{Q}}(x_0)\}$, $\Re\{\sur{\mathfrak{P}}
(x_0)\}$ and $\Im\{\sur{\mathfrak{P}}(x_0)\}$ (together with \eqref{Ws} 
applied at $x_0$) provide three relations for $\eta(x_0)$, $\eta_x(x_0)$ 
and $\omega$ (relations not written {\em in extenso\/} here, for brevity).


\section{Reconstruction procedure}\label{secnumrec}

From measurements of the bottom pressure $\bot{p}$, the free surface 
reconstruction procedure takes the following form. The first step is 
to choose a suitable basis of functions, generally a Fourier polynomial 
or elliptic functions \cite{Clamond2013,ClamondConstantin2013}. The 
best choice is the one providing the best fit among data with a minimum 
of eigenfunctions. Here, we consider only Fourier polynomials for simplicity, 
but the procedure is identical for any basis of functions expressible 
in the complex plane.
Thus, the wavenumber $k$ and the coefficients $\mathfrak{p}_n$ of the 
$N$th order Fourier polynomials $\bot{p} \approx \sum_{n=-N}^N 
\mathfrak{p}_n \ue^{\ui nkx}$, can be determined by least-square minimisation 
\cite{ClamondBarthelemy1995}. From \eqref{defdepth}, we know that 
$\mathfrak{p}_0 = g\depth$ and, since the acceleration due to gravity $g$ 
is known, this relation gives an expression for the mean water depth $\depth$. 
Thus, after this first step, parameters $g$, $d$, $k$ and $\mathfrak{p}_n$ 
are known explicitly and the bottom pressure can be extended everywhere in 
the bulk of the fluid.

From the analytic approximation of $\bot{p}$, the holomorphic pressure 
$\mathfrak{P}$ is, by definition, 
\begin{equation}
\mathfrak{P}(z) \approx \sum_{n=-N}^N \mathfrak{p}_n \ue^{\ui n k (z + 
\ui d)} = \sum_{n=-N}^N \mathfrak{p}_n \ue^{- n k \depth} \ue^{\ui n k z}, 
\end{equation}
so one obtains at once
\begin{equation}
\sur{\mathfrak{Q}}(x) = \int_{0}^{x} \left[\/ \sur{\mathfrak{P}}(x') - 
g\depth\/\right] \ud x' \approx \sum_{|n|>0}^N \frac{\ui \mathfrak{p}_n}
{n k} \frac{\ue^{-nka} - \ue^{\ui nk(x+\ui\eta)}}{\ue^{-nkd}} .
\end{equation}
After this second step, $\sur{\mathfrak{P}}$ and $\sur{\mathfrak{Q}}$ 
are known explicitly as functions of $\eta$. The wave surface profile 
$\eta$ is then determined solving the algebraic (i.e. not differential, 
nor integral) equation \eqref{Qseta} which, in general, can only be 
achieved numerically. This results in $\eta$ being obtained as a 
function of the parameters $a$, $b$, $\omega$ and $\sur{B}$. These 
parameters are then determined solving the nonlinear equations 
\rf{Bhat}, \rf{Pa}, \rf{Pb} and the ones at $x_0$. 


\begin{figure}
\centering

\subfloat[]{
\includegraphics*[width=.95\textwidth]{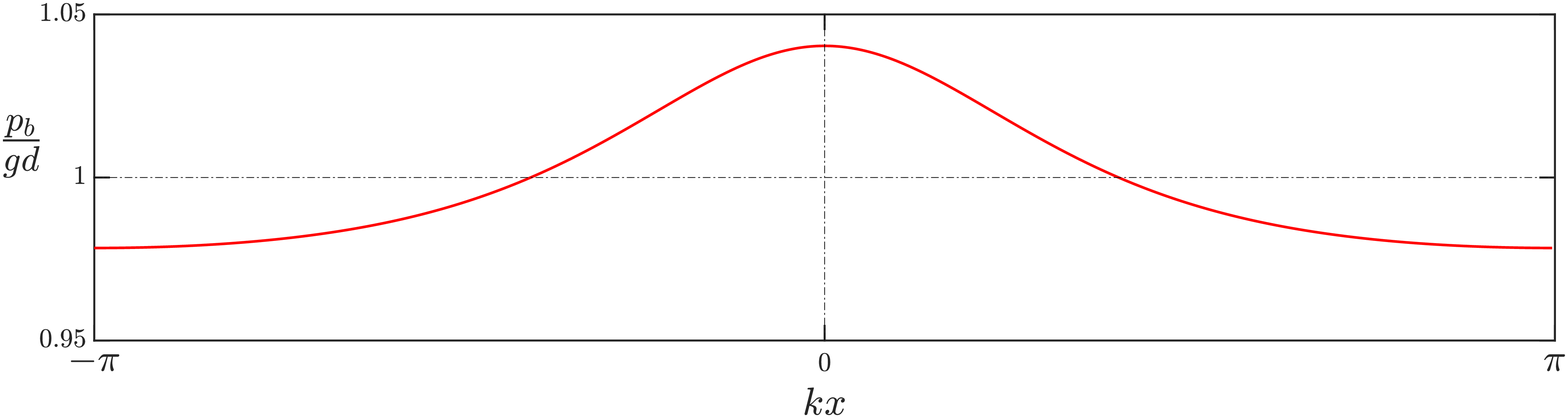}\label{Fig1a}
} \\ \vspace{.2em} 
\subfloat[]{
\includegraphics*[width=.95\textwidth]{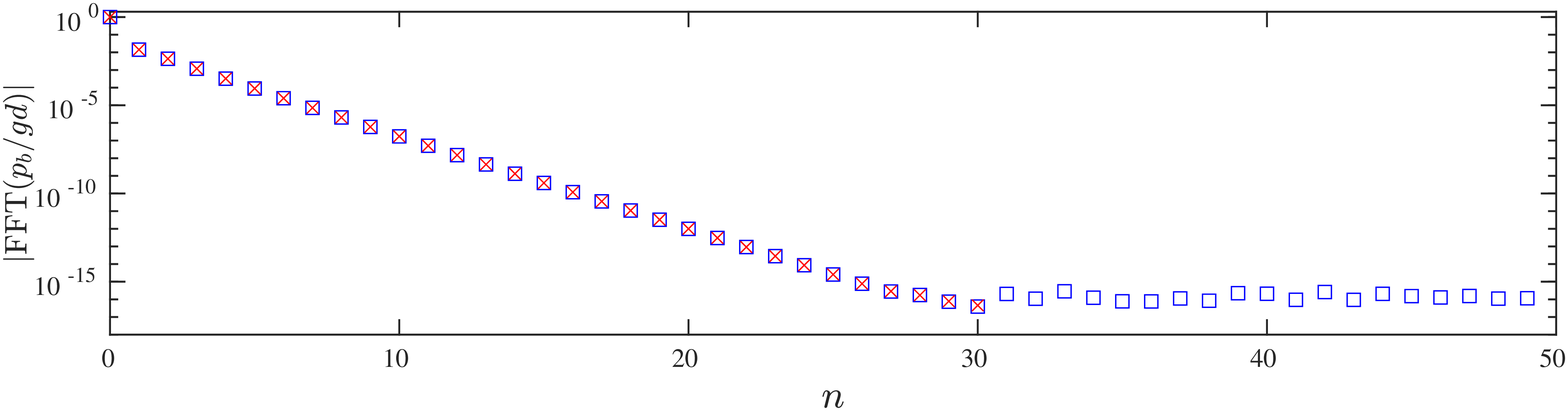}\label{Fig1b}
} \\ \vspace{.2em}
\subfloat[]{
\includegraphics*[width=.95\textwidth]{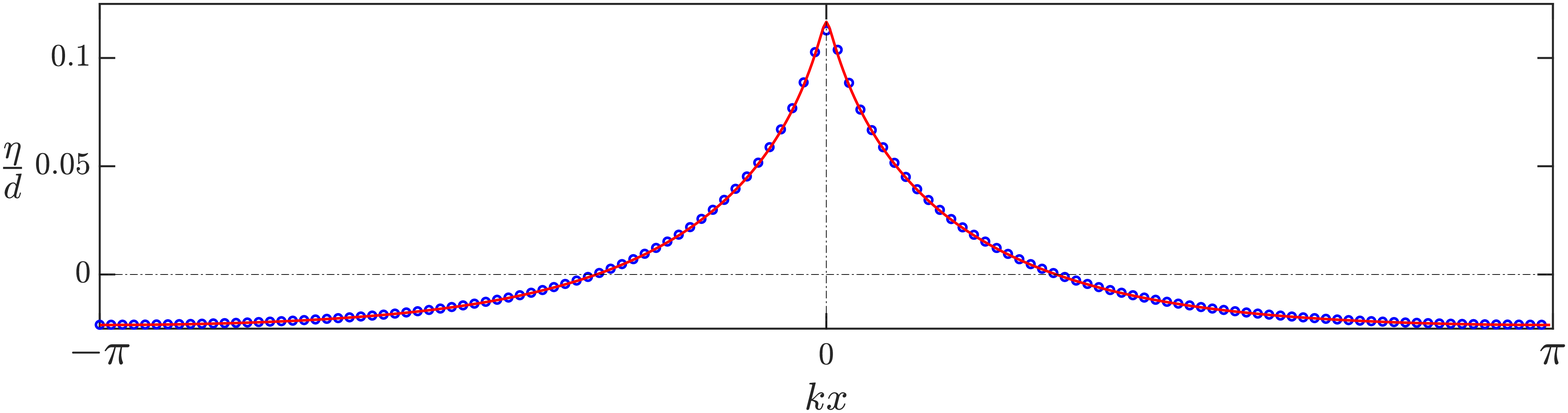}\label{Fig1c}
} 

\caption{Example of surface recovery for $\omega\sqrt{d/g}=-1.7$, $H/d=0.14$ 
and $L/\depth=2\pi$: (a) pressure at the bottom; (b) Fourier 
spectrum extracted from bottom pressure (blue squares) and fitted harmonics used 
for the reconstruction (red crosses); (c) recovered wave profile (red line) versus 
exact surface profile (blue circles).}
\label{Fig1}
\end{figure}

The numerical procedure consists in solving simultaneously the 
nonlinear set of equations \rf{Qseta}, \rf{Bhat}, \rf{Pa}, \rf{Pb} 
(and the ones at $x_0$) to recover the surface wave profile and related 
parameters. To this end, we use an iterative root-finding algorithm 
(the built-in function \textsf{fsolve} in \textsc{Matlab}) supplemented 
with initial values given by the linear theory \cite{KishidaSobey1988}.
When needed, $\eta_x$ is computed directly from its explicit expression 
\rf{deta}.

\section{Example}

In order to validate our procedure, we computed exact rotational 
waves with an accurate numerical algorithm similar to that of 
\cite{DaSilvaPeregrine1988}. We then obtained bottom pressures 
that are considered the `measured' ones from which the free surface 
and vorticity are recovered. Throughout the procedure, we monitor 
the convergence of our iterative algorithm using a tolerance 
of $10^{-12}$. When converged, the numerical 
solution is compared with the exact surface wave profile 
$\eta^{\textrm{ex}}$ and vorticity $\omega^{\textrm{ex}}$.

For waves of relatively small amplitudes, as well as small 
$\omega$, we witness a rapid convergence of the recovery procedure 
described above. The numerical errors of the recovered surface 
and vorticity are, respectively, $\epsilon_{\eta}=\lVert \eta 
- \eta^{\textrm{ex}} \rVert_{\infty}<10^{-8}$ and 
$\epsilon_{\omega}=\lvert \omega - \omega^{\textrm{ex}} \rvert 
< 10^{-6}$.
This excellent agreement is reached with as few as $N=5$ harmonics 
when considering the pressure Fourier expansion. However, for steady 
rotational waves with a steeper profile (thus departing from linear 
theory), the procedure naturally requires a larger number of harmonics.

\begin{figure}
\centering

\subfloat[]{
\includegraphics*[width=.95\textwidth]{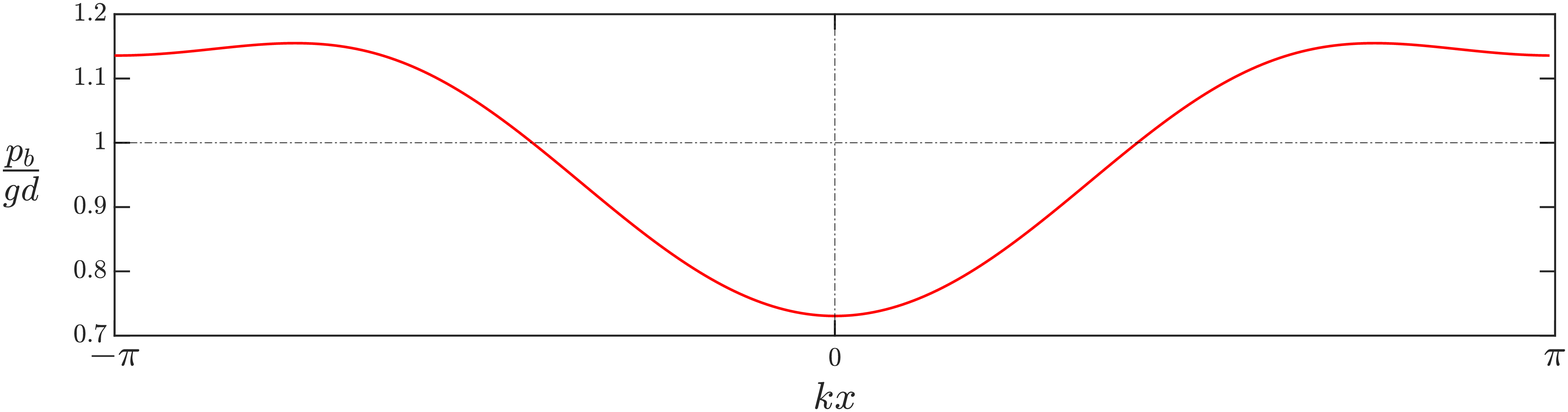}\label{Fig2a}
} \\ \vspace{.2em} 
\subfloat[]{
\includegraphics*[width=.95\textwidth]{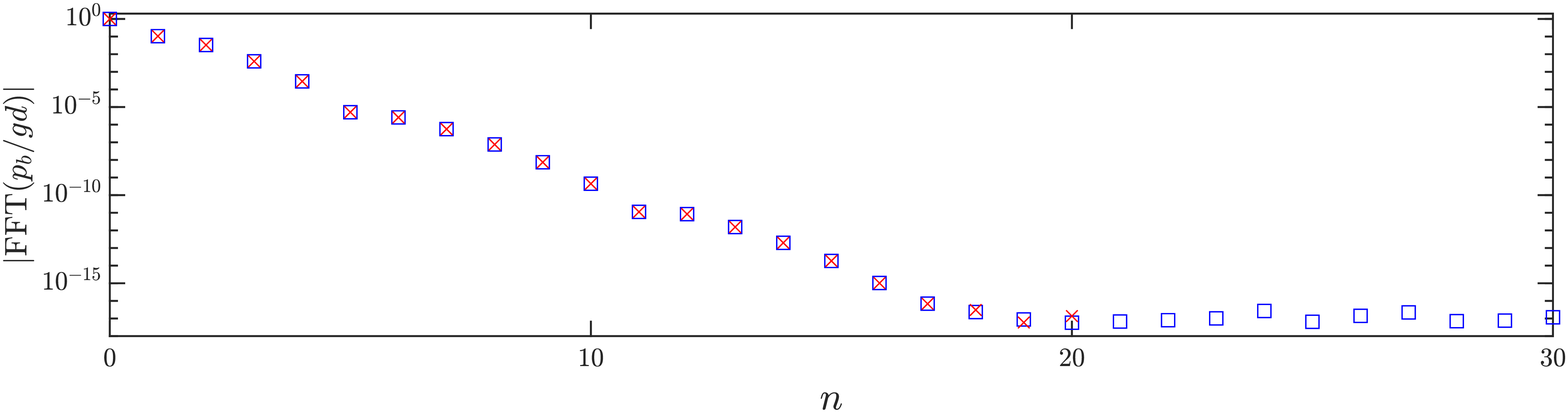}\label{Fig2b}
} \\ \vspace{.2em}
\subfloat[]{
\includegraphics*[width=.95\textwidth]{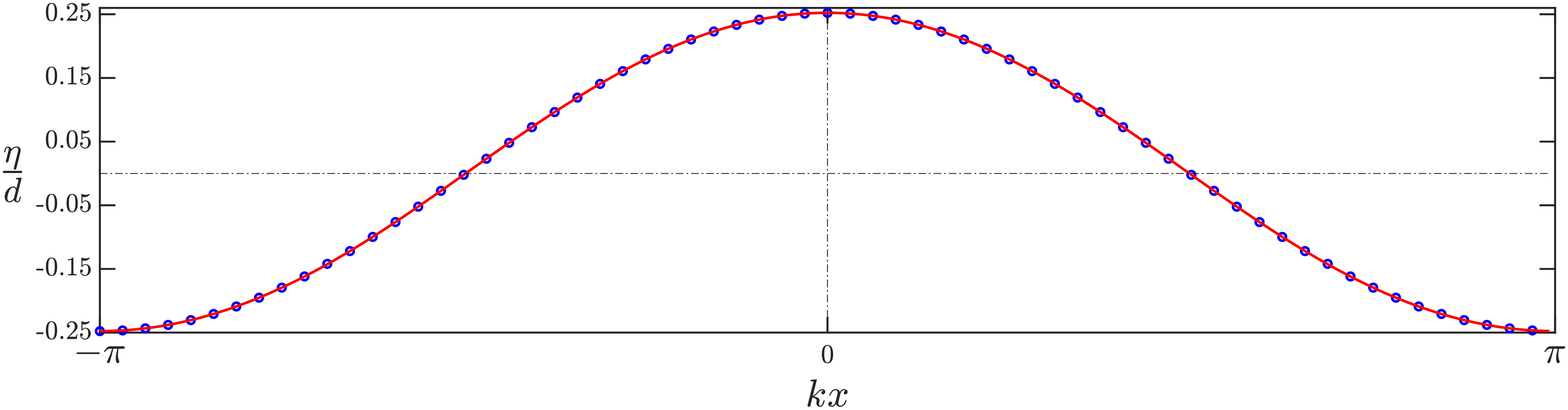}\label{Fig2c}
} 

\caption{Same as Figure \ref{Fig1} for the case $\omega\sqrt{\depth/g}=3$, 
$H/d=0.5$ and $L/\depth=2\pi$.}\label{Fig2}
\end{figure}

Here, we illustrate our surface wave recovery procedure with 
numerical examples depicted in Figures \ref{Fig1} and \ref{Fig2} 
for a domain of size $L/d=2\pi$ (rather deep water for which 
surface recovery is {\em a priori\/} difficult). For our first 
example, we examine a steep wave with negative vorticity 
$\omega\sqrt{d/g}=-1.7$.
Although the steep wave in figure \ref{Fig1c} would appear to be 
quite challenging to compute, we still recover the correct surface 
profile using $N=30$ harmonics in our Fourier expansion (c.f. panel 
\ref{Fig1b}), with the recovered solutions showing an excellent 
agreement with the exact data ($\epsilon_{\eta} \approx 10^{-4}$ 
and $\epsilon_{\omega} \approx 10^{-4}$).

For the second case of interest (Figure \ref{Fig2}), we take the positive vorticity 
$\omega\sqrt{\depth/g}=3$. This value is greater than that predicated by linear wave 
theory for the existence of critical layers. Indeed, linear waves with $c_0>0$ never 
allow for stagnation points when  $\omega<0$, while they contain stagnation points for 
$\omega>0$ iff $\tanh(kd)/kd \leq \omega^2 d /(g+\omega^2 d)$ 
\cite{ConstantinVarvaruca2011,DaSilvaPeregrine1988}.
This example is of special interest because it involves three stagnation 
points per wavelength: two at the bottom (about half way between crests and troughs) 
and one within the fluid (under the crests), c.f. Figure 14a of 
\cite{DaSilvaPeregrine1988}. The surface recovery still works relatively well in this case;  
taking $N=20$ Fourier modes,  errors are $\epsilon_{\eta} \approx 10^{-3}$ 
and $\epsilon_{\omega} \approx 10^{-4}$ (see Figure \ref{Fig2}). 

We observe from panel \ref{Fig1a} that the bottom pressure distribution for our choice of wave configuration 
displays a monotonic increase between trough and crest, which is not matched by that illustrated in panel 
\ref{Fig2a}. This is not an artefact solely of the difference in signs of the vorticities, but rather also their magnitudes --- an examination of explicit linear solutions for water waves with constant vorticity \cite{BrinkKjaer1976}
illustrates the richness in behaviour of the pressure fluctuations even for small amplitude waves, both with regard to its monotonicity properties, and the location of its 
extrema. For nonlinear waves, the precise qualitative behaviour of the bottom pressure fluctuations due to wave motion (and the location of its extrema) was only recently rigorously established for
irrotational  periodic waves  in \cite{Constantin2016}, 
and there are presently no similarly rigorous mathematical results for nonlinear waves with constant vorticity.  It is clear from panels \ref{Fig1a} and \ref{Fig2a} 
that considerable insight into this matter can be gained from a numerical approach.

\section{Discussion}
In this paper, we have presented a procedure for recovering 
fully nonlinear wave 
surface profiles from bottom pressure measurements for flows with constant 
vorticity. The theoretical basis for this procedure involves a reformulation 
in terms of holomorphic functions $\mathfrak{P}$ and $\mathfrak{Q}$, 
respectively  introduced in \cite{ClamondConstantin2013} and 
\cite{Clamond2013}, followed by a numerical implementation scheme. 
We have demonstrated the efficiency of this approach for different flow  
regimes, including one which presents stagnation points in the fluid body 
(an archetypical feature of waves with constant vorticity). For future work, 
it would be interesting to expand our approach to incorporate variable 
vorticity. Additionally, the question remains open as to whether our approach 
can be adapted to recover waves which exhibit overhanging profiles. \\

\noindent{\bf Funding.} Joris Labarbe has been supported by the French government, 
through the $\mbox{UCA}^{\mbox{\tiny JEDI}}$ {\em Investments in the Future\/} 
project managed by the National Research Agency (ANR) with the reference number 
ANR-15-IDEX-01. David Henry acknowledges the support of the Science Foundation 
Ireland (SFI) under the grant SFI 21/FFP-A/9150. \\

\noindent{\bf Declaration of interests.} The authors report no conflict of interest.

\bibliographystyle{amsplain}
\bibliography{Biblio}

\providecommand{\bysame}{\leavevmode\hbox to3em{\hrulefill}\thinspace}
\providecommand{\MR}{\relax\ifhmode\unskip\space\fi MR }
\providecommand{\MRhref}[2]{%
  \href{http://www.ams.org/mathscinet-getitem?mr=#1}{#2}
}
\providecommand{\href}[2]{#2}
\begin{thebibliography}{10}

\bibitem{Aris1962}
R.~Aris, \emph{Vectors, tensors and basic equations of fluid mechanics}, Dover,
  1962.

\bibitem{BrinkKjaer1976}
O.~Brink-Kj{\ae}r, \emph{Gravity waves on a current: The influence of
  vorticity, a sloping bed, and dissipation}, Institute of Hydrodynamics and
  Hydraulic Engineering, (ISVA), no.~12, Technical University of Denmark, 1976,
  pp.~1--137.

\bibitem{Clamond2013}
D.~Clamond, \emph{{New exact relations for easy recovery of steady wave
  profiles from bottom pressure measurements}}, J. Fluid Mech. \textbf{726}
  (2013), 547--558.

\bibitem{Clamond2017}
\bysame, \emph{Remarks on {B}ernoulli constants, gauge conditions and phase
  velocities in the context of water waves}, App. Math. Lett. \textbf{74}
  (2017), 114--120.

\bibitem{Clamond2018}
\bysame, \emph{New exact relations for steady irrotational two-dimensional
  gravity and capillary surface waves}, Phil. Trans. R. Soc. A \textbf{376}
  (2018), no.~2111, 20170220.

\bibitem{ClamondBarthelemy1995}
D.~Clamond and E.~Barth{\'{e}}l{\'{e}}my, \emph{Experimental determination of
  the phase shift in the {S}tokes wave -- solitary wave interaction}, C. R. Ac.
  Sci. Paris IIb \textbf{320} (1995), no.~6, 277--280.

\bibitem{ClamondConstantin2013}
D.~Clamond and A.~Constantin, \emph{Recovery of steady periodic wave profiles
  from pressure measurements at the bed}, J. Fluid Mech. \textbf{714} (2013),
  463--475.

\bibitem{ClamondDutykh2018b}
D.~Clamond and D.~Dutykh, \emph{Accurate fast computation of steady
  two-dimensional surface gravity waves in arbitrary depth}, J. Fluid Mech.
  \textbf{844} (2018), 491--518.

\bibitem{ClamondHenry2020}
D.~Clamond and D.~Henry, \emph{Extreme water-wave profile recovery from
  pressure measurements at the seabed}, J. Fluid Mech. \textbf{903} (2020), R3,
  12.

\bibitem{Constantin2011}
A.~Constantin, \emph{Nonlinear water waves with applications to wave-current
  interactions and tsunamis}, CBMS-NSF Regional Conference Series in Applied
  Mathematics, vol.~81, SIAM, Philadelphia, PA, 2011.

\bibitem{Constantin2012}
\bysame, \emph{On the recovery of solitary wave profiles from pressure
  measurements}, J. Fluid Mech. \textbf{699} (2012), 376--384.

\bibitem{Constantin2016}
\bysame, \emph{Extrema of the dynamic pressure in an irrotational regular wave
  train}, Physics of Fluids \textbf{28} (2016), no.~11, 113604.

\bibitem{ConstantinStraussVarvaruca2016}
A.~Constantin, W.~Strauss, and E.~Varvaruca, \emph{Global bifurcation of steady
  gravity water waves with critical layers}, Acta Math. \textbf{217} (2016),
  no.~2, 195--262.

\bibitem{ConstantinStraussVarvaruca2021}
\bysame, \emph{Large-amplitude steady downstream water waves}, Comm. Math.
  Phys. \textbf{387} (2021), no.~1, 237--266.

\bibitem{ConstantinVarvaruca2011}
A.~Constantin and E.~Varvaruca, \emph{Steady periodic water waves with constant
  vorticity: regularity and local bifurcation}, Arch. Ration. Mech. Anal.
  \textbf{199} (2011), no.~1, 33--67.

\bibitem{DaSilvaPeregrine1988}
A.~T. Da~Silva and D.~H. Peregrine, \emph{Steep, steady surface waves on water
  of finite depth with constant vorticity}, J. Fluid Mech. \textbf{195} (1988),
  281--302.

\bibitem{Henry2013}
D.~Henry, \emph{On the pressure transfer function for solitary water waves with
  vorticity}, Math. Ann. \textbf{357} (2013), no.~1, 23--30.

\bibitem{HenryThomas2018}
D.~Henry and G.~P. Thomas, \emph{Prediction of the free-surface elevation for
  rotational water waves using the recovery of pressure at the bed}, Philos.
  Trans. Roy. Soc. A \textbf{376} (2018), no.~2111, 20170102, 21.

\bibitem{KishidaSobey1988}
N.~Kishida and J.~Sobey, \emph{Stokes theory for waves on linear shear
  current}, J. Engin. Mech. \textbf{114} (1988), no.~8, 1317--1334.

\bibitem{OkamotoShoji2001}
H.~Okamoto and M.~Shoji, \emph{The mathematical theory of permanent progressive
  water-waves}, Advanced Series in Nonlinear Dynamics, vol.~20, World
  Scientific Publishing Co., Inc., River Edge, NJ, 2001.

\bibitem{OVDH}
K.~L. Oliveras, V.~Vasan, B.~Deconinck, and D.~Henderson, \emph{Recovering the
  water-wave profile from pressure measurements}, SIAM J. Appl. Math.
  \textbf{72} (2012), no.~3, 897--918.

\bibitem{ThomasKlopman1997}
G.P. Thomas and G.~Klopman, \emph{Wave-current interactions in the nearshore
  region}, Gravity waves in water of finite depth (J.~N. Hunt, ed.), Advances
  in Fluid Mechanics, Computational Mechanics Publications, 1997, pp.~255--319.

\end{thebibliography}

\end{document}